\documentclass[a4paper,11pt]{article}
\pdfoutput=1 

\usepackage{jcappub} 

\usepackage[T1]{fontenc} 
\usepackage[normalem]{ulem}
\usepackage{amsmath}
\usepackage{amssymb}
\usepackage{geometry} 
\usepackage{booktabs} 
\usepackage{array} 
\usepackage{paralist} 
\usepackage{verbatim} 
\usepackage{graphicx}
\usepackage[backend=biber,style=nature]{biblatex}
\usepackage[utf8]{inputenc}
\usepackage{multirow}
\usepackage[dvipsnames*,svgnames]{xcolor}

\usepackage{wasysym} 
\usepackage{subfigure} 
\usepackage{enumitem}
\usepackage{lineno}

\title{\boldmath 
Present and Future constraints on Secluded Dark Matter in the Galactic Halo with TeV Gamma-ray Observatories}

\author[a]{Guilherme N. Fortes,}
\author[b,c,d]{Farinaldo S. Queiroz,}
\author[a,b]{Clarissa Siqueira,}
\author[a]{Aion Viana}

\affiliation[a]{Instituto de F\'isica de S\~ao Carlos, Universidade de S\~ao Paulo, Av. Trabalhador S\~ao-carlense 400, S\~ao Carlos-SP, 13566-590, Brasil.}
\affiliation[b]{International Institute of Physics, Federal University of Rio Grande do Norte, Campus Universitário, Lagoa Nova, Natal-RN 59078-970, Brazil}
\affiliation[c]{Departamento de F\'isica, Universidade Federal do Rio Grande do Norte, 59078-970, Natal,
RN, Brasil}
\affiliation[d]{Millennium Institute for SubAtomic Physics at the High-energy frontier, SAPHIR, Chile}

\emailAdd{aion.viana@ifsc.usp.br}
\emailAdd{guilherme.fortes@usp.br}
\emailAdd{farinaldo.queiroz@ufrn.br}
\emailAdd{csiqueira@ifsc.usp.br}

\abstract{The dark matter relic density may be governed by the presence of new mediators that connect the dark matter field with the Standard Model particles. When the dark matter particle mass is larger than the mediator's, the pair production of mediators is kinematically open. This setup is known in the literature as secluded dark matter. Motivated by the appearance of secluded dark matter in several model building endeavours, we investigate the sensitivity of TeV gamma-ray instruments in the Southern Hemisphere namely, H.E.S.S., CTA, and SWGO to secluded dark matter annihilating in the Galactic Halo. We exploit the complementarity aspects of these detectors to find restrictive bounds on the annihilation cross-section for different annihilation channels. In particular, for a dark matter particle mass of $2$~TeV, H.E.S.S. is able to constraint $\langle \sigma v \rangle \geq 4 \times 10^{-26}\,\, {\rm cm}^3\, {\rm s}^{-1}$ at 95\% confidence level for the $4q$ and $4\tau$ channel, while CTA will be sensitive to  $\langle \sigma v \rangle \geq 7 \times 10^{-27}\,\, {\rm cm}^3\, {\rm s}^{-1}$ and SWGO $\langle \sigma v \rangle \geq 6 \times 10^{-27}\,\, {\rm cm}^3\, {\rm s}^{-1}$ for the $4\tau$ channel, both well below the thermal relic cross-section. In fact, the combination of CTA and SWGO will be able to probe cross-sections below the thermal relic value for dark matter particles in the whole mass range between 100 GeV and 100 TeV in the $4q$ and $4\tau$ channels, and between 100 GeV and $\sim$40 TeV in the $4b$ channel.}

\begin{document}
\maketitle
\flushbottom

\section{Introduction} 
\hfill 

The presence of dark matter (DM) is required at small and large scales and throughout the Universe's history \cite{1991Natur.352..769P,1993ppc..book.....P}. Dwarf galaxies have a size of about $<\sim$~kpc and are notoriously DM dominated objects \cite{2008Natur.454.1096S}. The clusters of galaxies that have $\sim 10$~Mpc are DM dominated objects as well, as firstly noticed by F. Zwicky in 30s \cite{Zwicky:1933gu}, wisely evoking the virial theorem. The well-known Bullet Cluster event, which concerns the observation of a collision between galaxy clusters, clearly attested to the need for DM. Today, over 70 collisions between galaxy clusters have been observed, and collectively they detect the existence of dark mass at $7.6\sigma$ significance \cite{Harvey:2015hha}. Having in mind the data from Baryon Acoustic Oscillations \cite{Eisenstein:2005su}, Cosmic Background Radiation \cite{Ade:2015xua,Aylor:2017haa,Aghanim:2018eyx,Aiola:2020azj}, Big Bang Nucleosynthesis \cite{Zyla:2020zbs}, and structure formation \cite{Primack:1997av}, the presence of DM being the most abundant form of matter in our Universe is irrefutable. Although, despite the vast data available, we have not unveiled its nature yet.

The most popular and one of the most compelling explanations for DM is known as WIMPs (weakly interacting massive particles), which have a mass in the GeV to TeV range, and naturally reproduce the observed DM abundance, $\Omega h^2 =0.120 \pm 0.001$ \cite{Aghanim:2018eyx} (Planck), if they have an annihilation cross-section around the weak scale \cite{Arcadi:2017kky}. These WIMPs' characteristics are particularly interesting because they can produce detectable gamma-ray signals at our telescopes \cite{Abramowski:2011hh,Abramowski:2011hc,Ackermann:2015zua,Ahnen:2016qkx,Abdallah:2016ygi,Hoof:2018hyn,CTAConsortium:2018tzg,Li:2018kgy,Oakes:2019ywx,Viana:2019ucn,Abdallah:2020sas,Acharyya:2020sbj}. A gamma-ray signal from DM is proportional to the DM annihilation rate and density, but inversely proportional to the DM mass. Therefore, the search for gamma-ray signals from DM annihilation probes simultaneously particle physics and astrophysical quantities. One of the important ingredients in the search is the final state resulting from DM annihilation. In general, DM particles might annihilate into any of the Standard Model (SM) particles. Annihilations into light quarks and heavy lepton pairs are typically investigated as benchmark annihilation channels. Usually, the limits on the DM annihilation cross-section are reported in terms of these benchmark final states. However, there are many popular DM models that feature annihilation channels different from the benchmark ones, and in some cases, those are dominant. One clear example where the annihilation channels are quite different is those where DM particles appear in the final state, known as semi-annihilations \cite{Belanger:2012vp,Cai:2015tam,Arcadi:2017vis}. 

Typically, the DM particles interact with the SM ones via the presence of new particles, called mediators, which are often scalar or vector particles. These particles give rise to the so-called \textit{portals} that dictate the DM annihilation rate and thus the relic density. In general, the DM particle can be heavier than these mediators, and if that happens, the DM self-annihilation into pairs of these particles is kinematically open. This scenario where the DM particles annihilate mostly into mediators and not into SM particles was named \textit{secluded}. Notice that secluded DM can always occur as long as the DM particle is heavier than the mediator. We highlight that having secluded DM annihilations in a model does not mean that these secluded annihilations are dominant. It depends on the details of the model, but generally, when DM mass is much larger than the mediator's, secluded channels become more relevant, and therefore they should be the target of experimental scrutiny under gamma-ray telescopes. Besides these most common cases, there are models where the DM particle is either confined to the dark sector \cite{vonHarling:2012sz,Kim:2016csm}, or form bound states, where secluded annihilations are generally dominant \cite{Cirelli:2016rnw} and also can be probed by these gamma-ray observatories. It is important to emphasize that secluded scenarios offer a softer spectrum compared to direct annihilation, which could be used to distinguish possible signatures at the detectors.

Ground-based gamma-ray observatories consist of two kinds: \emph{i}) \textit{imaging atmospheric Cherenkov telescope} (IACT) arrays that measure the Cherenkov light in extensive air showers generated in the atmosphere by very-high-energy (VHE, E $\gtrsim$ 100 GeV) gamma rays; \emph{ii}) particle detectors that directly sample the particles in the extensive air showers when they reach the ground. Both these techniques allow the instruments to accurately reconstruct the direction and the energy of the primary gamma ray. Typically, IACTs have a better angular resolution and are sensitive to lower gamma-ray energies, on the other hand, they have a significantly smaller field-of-view and duty cycle with respect to air shower particle detectors~\cite{DENAUROIS2015610}.    

These gamma-ray observatories are particularly sensitive to DM searches at the TeV mass range, due to their large collection area and good angular resolutions \cite{Viana:2019ucn,Aguirre_Santaella_2020}. Among the several instruments around the world, those placed in the Southern Hemisphere have the advantage of having the Galactic Center (GC) transiting close to the zenith. The GC is arguably the most interesting region to look for a DM annihilation signal, as it is predicted to be the brightest source of gamma rays from this kind of interaction. Even considering possible signal contamination from other astrophysical sources, it is one of the most promising targets to detect the presence of these new massive particles. In this work, we will use data from the current H.E.S.S experiment \cite{HESS:2015cda,Abdallah:2016ygi}, and also the perspectives for the future observatories \textit{Cherenkov Telescope Array} (CTA)~\cite{CTAConsortium:2018tzg} and the \textit{Southern Wide field-of-view Gamma-ray Observatory} (SWGO).

Therefore, motivated by the presence of secluded annihilation in a variety of models, we will assess the sensitivity of current and future gamma-ray observatories in the Southern Hemisphere to secluded DM. In summary, the key aspects of our work are: 

i)  We consider the inner degree around the GC but also extend the searches to a large 10 degrees radius circular region of the Galactic Halo in this study; 

ii) We consider the latest H.E.S.S. results~\cite{Abdallah:2016ygi}, equivalent to 254h live-time exposure, and use the same dataset and instrument effective area as that study to derive limits on the annihilation cross-section of secluded DM; 

iii) We assess the CTA sensitivity to the channel of interest using the most up-to-date, publicly available instrument design and response functions; 

iv) We determine for the first time the SWGO potential to probe secluded DM, it is important to emphasize that we are considering just DM annihilation (we are not including decaying DM); 

v) We exploit the complementarity from these telescopes and highlight the instrumental difference between them and their consequential performance at probing DM masses ranging from $100$~GeV to $100$~TeV, extending previous searches to masses above $10$~TeV. 

Then, our paper is structured in the following way: in section \ref{sec: secluded}, we discuss the secluded model at TeV-scale and compute the gamma-ray spectra for each of our benchmarks; in section \ref{sec:observatory}, we provide a brief introduction about the observatories studied in this work; in section \ref{sec:methodology}, we address the methodology used to compute the limits; in section \ref{sec:results}, we present our results, and, finally, in section \ref{sec:conc}, we draw our conclusions.


\section{Secluded Dark Matter at TeV}
\label{sec: secluded}
\hfill 

\begin{figure}[!ht]
    \centering
    \includegraphics[width=0.4\textwidth]{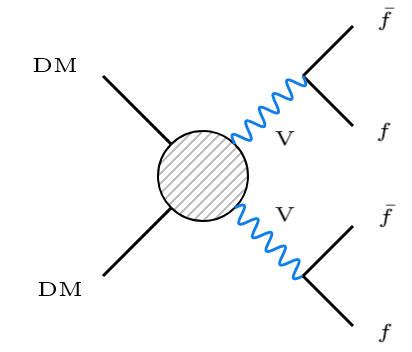}
    \caption{Feynman diagram representing the secluded DM setup we are probing, where DM particles annihilate into a pair of short-lived light states, which then decay into SM particles. The SM particles then give rise to gamma-rays via either hadronization processes or final state radiation.}
    \label{fig:diagram}
\end{figure}

Dark matter particles annihilate into SM particles, often through mediators. However, when the DM particles are sufficiently heavy they may annihilate dominantly into mediators (see the Feynman diagram in Fig.~\ref{fig:diagram}). This scenario is called secluded. Models, where the processes that set the relic density are the same as those that govern the DM-nucleon scattering, suffer from stringent bounds stemming from direct detection experiments \cite{Arcadi:2017kky}. Hence, secluded annihilations are, sometimes, evoked to break the connection between relic density and direct detection rates, and weaken the direct detection bounds. In other words, by making secluded annihilation dominant, one can relax the bounds on the parameter space of some models \cite{Pospelov:2007mp,Kim:2016csm,Dutra:2018gmv}. There are other interesting ways to successfully untie the annihilation rate to the scattering rate which are outside the scope of secluded DM, but we refer the reader to Refs.\cite{Mahanta:2019sfo,Chanda:2019xyl,Arcadi:2020aot}. We highlight that there are several secluded DM constructions in the literature, involving dark photons \cite{Pospelov:2007mp,Pospelov:2008jd,Batell:2009zp,DeAngelis:2017gra,Breitbach:2018ddu,Dutra:2018gmv}, heavy vector bosons \cite{Dedes:2009bk,Fortes:2015qka,Fortes:2017kca} even scalar fields \cite{Yaguna:2019cvp,PhysRevD.100.095020,Belanger:2020hyh} (see \cite{Okawa:2016wrr} for other possibilities). 

The interesting aspect of secluded DM as far as indirect DM detection is concerned is the change from $2\rightarrow 2$ annihilations to $2\rightarrow 4$. The energy spectrum (photons produced per annihilation) is modified in the presence of such $2\rightarrow 4$ processes.  In our study, we will assume that the mediator is light compared to the DM particle\footnote{But it is important to keep in mind that the mass of the mediator can affect the overall results if its mass is comparable to the DM mass \cite{Profumo:2017obk}.}. Also, we will explore certain simple benchmarks where the mediators are decaying into a single particle type, namely, into $V \to 2e$, $V \to 2\mu$, $V \to 2\tau$, $V \to 2q$ (with $q$ representing the light quarks), and $V \to 2b$.    

In this context, in order to compute the gamma-ray spectrum, since we are dealing with heavy DM (up to 100~TeV), in which some corrections are not included \cite{Ciafaloni:2010ti,Christiansen:2014kba} by the frequently used numerical packages like Pythia \cite{Sjostrand:2006za} and PPPC4DMID \cite{Cirelli:2010xx}, we computed the spectrum in the rest frame of the mediator and boosted it to the DM one, following \cite{Elor:2015bho} and, also, \cite{Berlin:2014pya,Elor:2015tva,Gao:2017pym}. So, in this case, the gamma-ray spectrum is given by,

\begin{equation}
    \frac{dN^\gamma}{dx_1} = 2 \int^{t_{1,max}}_{t_{1,min}} \frac{d x_0}{x_0 \sqrt{1-\epsilon_1^2}} \frac{dN^\gamma}{dx_0}
\end{equation}
where

\begin{eqnarray}
  t_{1,min} &=& \frac{2 x_1}{E_1^2} \left(1-\sqrt{1-\epsilon_1^2}\right) \\
  t_{1,max} &=& Min\left[1,\frac{2 x_1}{E_1^2} \left(1+\sqrt{1-\epsilon_1^2}\right)\right]
\end{eqnarray}
with $dN^\gamma/dx_0$ the gamma-ray spectrum in the mediator rest frame, the $x_0=2 E_0/m_V$, for $m_V$ being the mediator mass, and $E_0$ the energy at the rest frame of the mediator, and $x_1 = E_1/m_{DM}$ with $E_1$, the photon energy in the rest frame of the DM particle. The parameter $\epsilon_1=m_V/m_{DM}$, relates the masses of the mediator and the DM particle, and we can also define the following parameter, 

\begin{equation}
    \epsilon_f = \frac{2 m_f}{m_V},
\end{equation}
which provides a relation between the mass of the mediator with the SM particles in the decay products. Using the above relations, we can get

\begin{equation}
    m_{DM} = \frac{2 m_f}{\epsilon_f \epsilon_1}
\end{equation}
and it is possible to use the parameter $\epsilon_f$, which relates the mass of the mediator with the decay products, as a free parameter in our analysis as well as the annihilation channels.

\begin{figure}[!ht]
    \centering
    \includegraphics[width=0.49\textwidth]{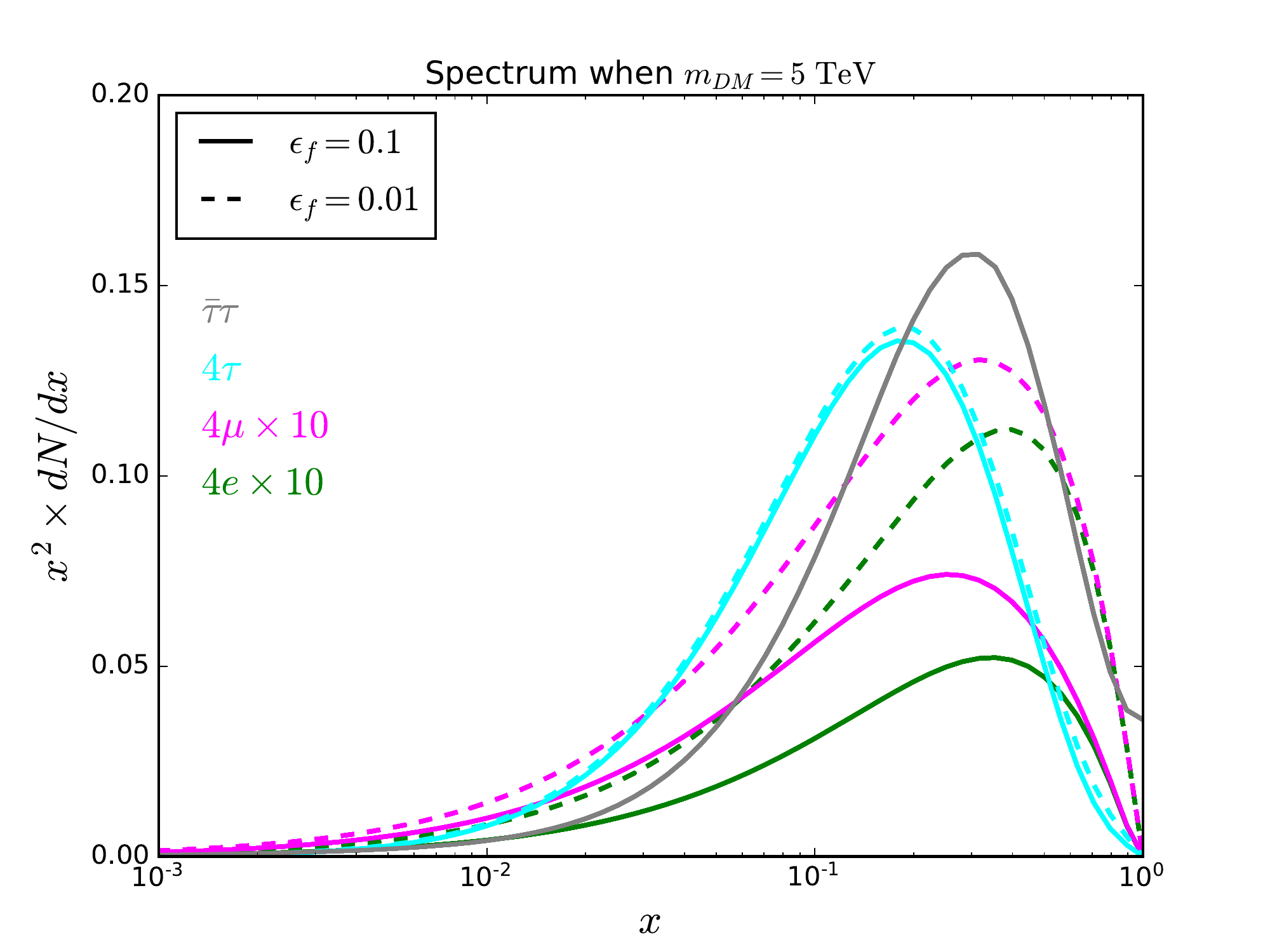}
    \includegraphics[width=0.49\textwidth]{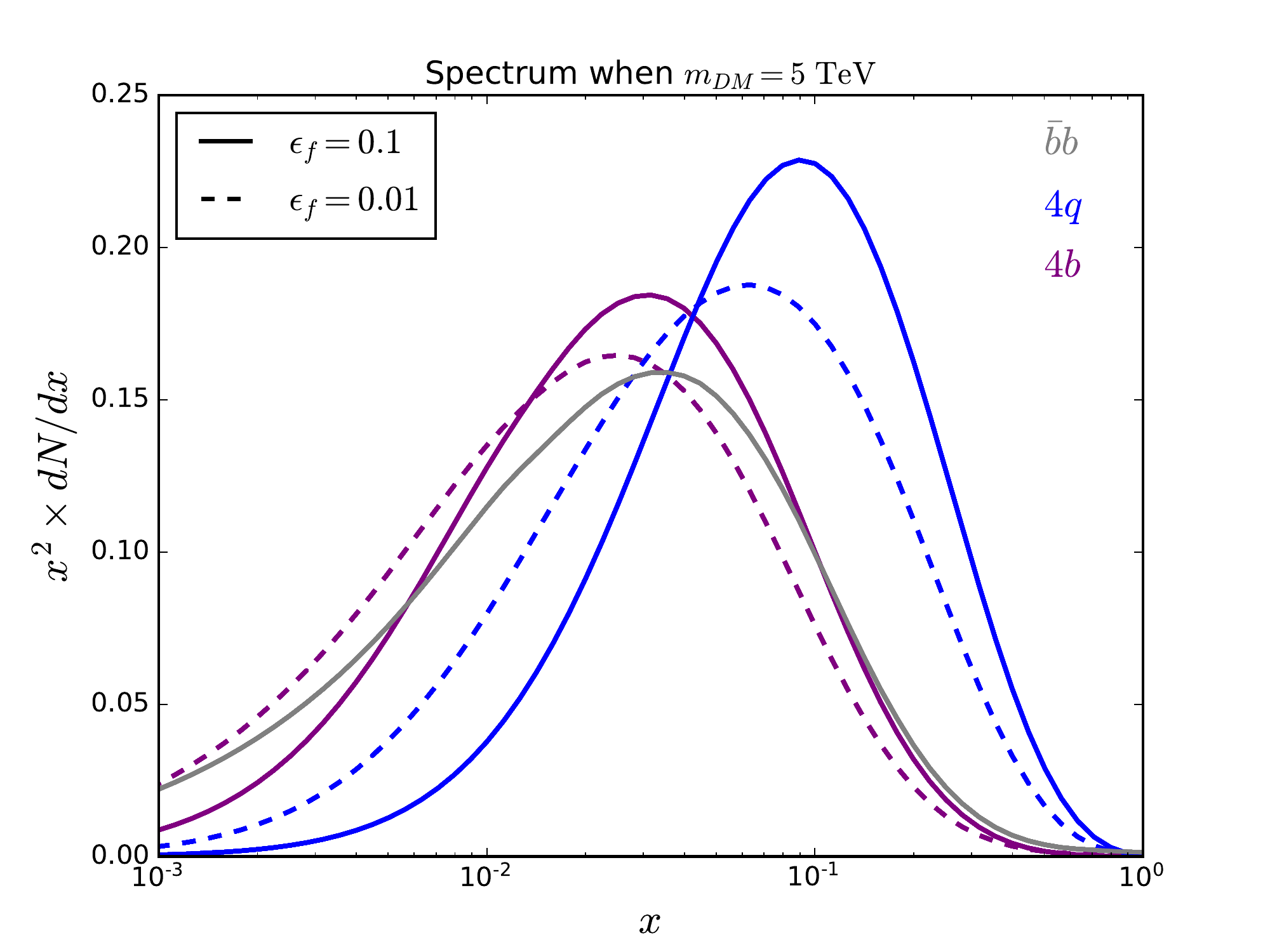}
    \caption{Energy spectrum expected for several annihilation channels studied here, between them, $V \to 2b$ (purple), $V \to 2q$ (blue) (left), $V\to 2\tau$ (cyan), $V \to 2\mu$ (magenta) and $V \to 2e$ (green) (right). We choose $\epsilon_f = 0.1$ ($0.01$) showed in continuous (dashed) lines, for the DM mass equal to $5$~TeV. For comparison, we include the spectrum for direct annihilation into $\bar{\tau}\tau$ (left) and $\bar{b}b$ (right).}
    \label{fig:dNdE}
\end{figure}

In Figure~\ref{fig:dNdE}, we present the $dN^\gamma/dE$ for the DM mass equal to $5$~TeV, for leptonic channels (left panel) and hadronic channels (right panel), for two values of the parameter $\epsilon_f = 0.1$ and $0.01$. The leptonic annihilation channels include $V\rightarrow2e$, $V\rightarrow2\mu$ and $V\rightarrow2\tau$, for comparison, we add the direct annihilation channel into $\bar{\tau}\tau$. While the hadronic annihilation channels are $V\rightarrow2b$, $V\rightarrow2q$, with $q$ representing light quarks, here we also include for comparison the direct annihilation into $\bar{b}b$. It is important to emphasize that we are computing the spectra for scalar mediators, however, this mediator can be a scalar or a vector, as demonstrated in \cite{Elor:2015tva}, since the difference in the spectrum is modest, so this analysis can be qualitatively regarded as model-independent. With the spectrum at hand, we can follow to the next section, where we will discuss the gamma-ray observatories addressed in this analysis. 

\section{Very-high-energy gamma-ray observatories in the Southern Hemisphere}
\label{sec:observatory}

\subsection{H.E.S.S. observations of the GC region}
\hfill 

The only gamma-ray observatory currently taking data in the Southern Hemisphere is H.E.S.S.\footnote{\url{www.mpi-hd.mpg.de/hfm/HESS/}}, an IACT array located in the Khomas Highland of Namibia at 23$^{\circ}$ South of latitude. H.E.S.S. has been monitoring the GC region for the past 16 years and has provided the most constraining limits to DM annihilation cross-sections of WIMPs at the multi-TeV mass range \cite{HESS:2015cda,Abdallah:2016ygi}. The deep observation campaign of the GC region undertaken by the H.E.S.S. telescope over the past decade has led to the detection of numerous astrophysical sources that shine in VHE gamma rays. Among them are H.E.S.S. J1745-290~\cite{Aharonian:2004wa}, a strong emission spatially coincident with the supermassive black hole Sagittarius A*, the supernova/pulsar wind nebula G09+01~\cite{Aharonian:2005br}, the supernova remnant H.E.S.S. J1745-303~\cite{Aharonian:2005kn}, and a diffuse emission extending along the Galactic plane~\cite{Aharonian:2006au, Abramowski:2016mir, Abdalla:2017xja}. 

The dataset obtained from 10 years of observations using the 4-telescope array of the H.E.S.S. phase-I instrument has been used to look for continuum~\cite{Abdallah:2016ygi} and line signals~\cite{Abdallah:2018qtu} from DM annihilations. Standard analyses of H.E.S.S. phase-I observations of the GC region provided about 250 hours of live time in the inner 1$^{\circ}$ of the GC, with a mean zenith angle of about 20$^{\circ}$ that yields an energy threshold of $160$~GeV.  In order to avoid contamination from the standard astrophysical background in the regions-of-interest (ROIs) for DM searches, a region of $\pm 0.3^{\circ}$ in Galactic latitude along the Galactic plane has been excluded from the dataset together with a disk of 0.4$^{\circ}$ radius centered at the position of J1745-303. A 2-dimensional likelihood ratio test statistics was used to look for any possible excess over the measured background.  

The background is measured using the reflected background method~\cite{Abdallah:2018qtu}. For each observation and each ROI, also called ON region, the background is determined from the region in the sky symmetric to it with respect to the pointing position, called OFF region. In this way, the ON and OFF regions have the same exposure, acceptance, angular size, and observational conditions, so no further correction is needed to compare both regions. No excess in the ON region with respect to background was found. In this work, we use the same ON and OFF event's energy distribution and instrument's effective area as it was used for the 10 years H.E.S.S. DM continuum searches~\cite{Abdallah:2016ygi}.

\subsection{CTA performance} 
\hfill 

One of the next generation of ground-based gamma-ray observatories will consist of the \textit{Cherenkov Telescope Array} (CTA\footnote{\url{www.cta-observatory.org/}}). CTA, currently under construction, will consist of two IACT arrays: one located in the Northern Hemisphere (La Palma, Canary Islands, Spain); the other situated in the Southern Hemisphere (Paranal, Chile). The experiment will contain telescopes of three different sizes: large, with a diameter of $23$m, medium, with $11.5$m and small, with $4$m. The different sizes and locations give to the CTA a large field-of-view and energy interval, with an energy sensitivity from $20$~GeV to $300$~TeV \cite{Acharyya:2020sbj}. The Southern array, CTA-South, placed at 24$^{\circ}$ South of latitude, will be able to observe the GC region with unprecedented accuracy. Because of that, a very deep survey of the GC region is planned as a Key Science Project of CTA with a large time commitment spread over several years at the beginning of its operation~\cite{CTAConsortium:2018tzg}.

The sensitivity to direct DM annihilation signals for CTA observations of the GC region, estimated recently in \cite{Acharyya:2020sbj}, shows that the CTA will be able to probe the standard annihilation cross-section at the TeV DM mass range for $W^+W^-$ and $\bar{b}b$ annihilation channels, a region never reached before. In this analysis, we will compute the sensitivity of the CTA for secluded annihilation channels, using the latest publicly-available instrument response functions (IRFs) of the ``Omega Configuration'' (formerly referred to as the \textit{baseline} configuration) of the CTA Southern site at average zenith angle 20$^{\circ}$ and optimized for 50 hours of observation\footnote{\url{www.cta-observatory.org/science/cta-performance/}}. 

In this analysis, we consider a circular $1^\circ$ region centered at the GC, and in order to minimize the background, we exclude the central region $|b|< 0.3^\circ$, with an observation time of $500$ hours. The main background for ground-based gamma-ray observatories consists of charged cosmic rays (CRs), both from hadronic (proton and nuclei) or leptonic (electron and positrons) origin, that have their air shower misidentified as being originated by a gamma-ray. Although there are several techniques that help separate CR showers from gamma rays, a large amount of residual background remains. The expected residual background determination for CTA has been performed through extensive Monte Carlo simulations~\cite{2019APh...111...35A}. Here, we use the so-called \textit{prod3b} version of the instrument response functions that include the residual background determinations.  

\subsection{SWGO model}
\hfill 

The other next-generation ground-based gamma-ray observatory is the  SWGO\footnote{\url{www.swgo.org}} \cite{Schoorlemmer2019}. SWGO is a wide field of view, and high-duty cycle, water Cherenkov particle detector array, planned to be built in South America at a latitude between 10 and 30 degrees South. Such an instrument will have unprecedented sensitivity in the multi-TeV energy scale, expected to reach from $500$~GeV to $2$~PeV, large FOV (45$^\circ$), and daily exposure of the GC with a good angular resolution ($<0.5^\circ$)~\cite{SGSO_WP}. It is currently in its research and development (\textit{R\&D}) phase. 

The SWGO shows an unprecedented sensitivity to the search for multi-TeV DM masses, as shown in \cite{Viana:2019ucn}, for example, the SWGO will be able to probe the standard annihilation cross-section for the $\bar{\tau}\tau$ direct annihilation channel until $m_{DM}=100$~TeV. And the most interesting point, the SWGO will be complementary to the CTA DM searches.   

In this work, for secluded DM annihilation, we consider a $10^\circ$ region centered at the GC, excluding the central $\pm 0.3^\circ$ of the Galactic latitude, with an observation time of $10$ years. We also use publicly-available\footnote{\url{https://github.com/harmscho/SGSOSensitivity}} instrument response functions. These IRFs have been produced for the science case studies presented in Ref.~\cite{SGSO_WP} and were used for the sensitivity estimates to DM annihilation of typical WIMP particles in the GC halo mentioned above (see Ref. \cite{Viana:2019ucn} for more details). The detector model is basically a scaled-up version of the current generation of air shower particle detectors, like High Altitude Water Cherenkov gamma-ray observatory (HAWC) \cite{HAWC_CRAB}, placed at a higher altitude in order to increase the sensitivity to sub-TeV gamma-rays. The ground coverage was also increased to 80\%, which is between the ground coverage of HAWC (57\%) and LHAASO ($\sim$100\%). The elevation is chosen at 5\,km altitude, which is close to the altitude of several site candidates currently under consideration by the SWGO collaboration. In addition, the background also consists of the identification of CR as gamma-rays, as in CTA.  
    
\section{Gamma-ray Fluxes from Secluded Dark Matter towards the Galactic Center} 
\label{sec:methodology}
\subsection{Annihilation of Dark Matter Particles}
\hfill 

The differential gamma-ray flux resulting from DM annihilation can be calculated by

\begin{equation}\label{gamma_flux}
\frac{d\phi_\gamma}{dE_\gamma}(E_\gamma, \Delta\Omega) = \frac{\langle\sigma v\rangle}{8\pi m^2_{DM}}\frac{dN}{dE_\gamma}\times J(\Delta\Omega).
\end{equation}

The first term corresponds to particle physics, relating the nature of DM properties and their interactions. $\langle\sigma$v$\rangle$ is the thermal-averaged velocity-weighted cross-section, $m_{DM}$ is the mass of DM and dN/$dE_\gamma$ is the annihilation differential spectrum into gamma rays. The second, astrophysical term, named J-factor, contains information on how DM is distributed across the galaxy and in which direction the observation is being made. More explicitly, it can be given by

\begin{equation}\label{j_factor_integral}
J(\Delta\Omega) = \int_{\Delta\Omega}\int_{l.o.s} d\Omega ds \: \rho^2_{DM}(r(s, \theta)),
\end{equation}
being the integral of the square of the density $\rho_{DM}$ along a line of sight (l.o.s) over a solid angle $\Delta\Omega$. The density function is conveniently parameterized as a function of r, between the observer and the target, with r becoming

\begin{equation}\label{r_parametrization}
r = \sqrt{s^2 + r_\odot^2 - 2sr_\odot \cos{\theta}} \, ,
\end{equation}
in which $s$ is the distance along the l.o.s, $r_\odot$ is the distance of the Sun to the GC, here taken to be $8.5$~kpc, and $\theta$ is the angle between the observation direction and the GC. 

\subsection{Galactic Halo density profiles and regions of interest}
\hfill 

The galactic DM density distribution of the Milky Way is not entirely known, containing sizable uncertainties, they come from the observations that are used to infer the dynamics of the Galaxy \cite{Benito:2019ngh,Benito:2020lgu}. Even though this determination of the DM density cannot be applied to the region of the GC, because there is a lack of dynamical information within the bulge region ($r<3$kpc), any profile adopted relies on the extrapolation of what was obtained at higher radii \cite{2019JCAP...09..046K,2020MNRAS.494.4291C}. Therefore there is freedom to choose any DM profile for the GC region \cite{2017PDU....15...90I}. In this paper, we have chosen a cuspy Einasto profile with parameters taken from Refs. \cite{Abdallah:2016ygi, Abramowski:2011hc}, to consistently compare to previously DM searches performed by H.E.S.S.~\cite{Abdallah:2016ygi}, CTA~\cite{CTA_ScienceTDR,Acharyya:2020sbj} and SWGO~\cite{Viana:2019ucn}\footnote{The effects of a cored profile to the sensitivity of both SWGO and CTA was estimated in Ref.~\cite{Viana:2019ucn}, where, by assuming a Burkert profile~\cite{Burkert:1995yz}, CTA limits would degrade by a factor of $\sim$166, whereas SWGO limits would get weaker by a factor of $\sim$48.}. 

    
The Einasto profile has the following parameterization
    
\begin{equation}\label{einasto}
    \rho_{Ein}(r) = \rho_s \, {\rm exp}\left(-\frac{2}{\alpha}\left[\left(\frac{r}{r_s}\right)^{\alpha}-1\right]\right),
\end{equation}
with $\rho_s$ and $r_s$ are the radius and density at which the logarithmic slope of the density is -2, respectively, and the parameter $\alpha$ controls the curvature of the profile.

The profile considered here has parameters chosen in accordance with Ref.\cite{Abdallah:2016ygi,Abdallah:2016ygi}, more explicitly, $\rho_s$ = 0.079, $r_s$ = 20 kpc, $\alpha = 0.17$. The parameter $\rho_s$ was calculated by assuming a local DM density of 0.39 GeV/cm$^3$ \cite{Catena:2009mf,2011MNRAS.414.2446M,Benito:2019ngh,2019JCAP...09..046K}.

For both H.E.S.S. and CTA, a circular 1$^{\circ}$ region centered at the GC was considered for the analysis. Following the H.E.S.S. strategy, this area was further subdivided into seven concentric rings as our spatial ROIs, all having a 0.1$^{\circ}$ width, starting at 0.3$^{\circ}$ from the GC. Additionally, as already mentioned, to minimize the various background gamma-ray sources located at the GC region, a $\pm$0.3$^{\circ}$ band in Galactic latitude was excluded from the considered ROIs.  In the case of SWGO, as in Ref.\cite{Viana:2019ucn}, we focus the analysis on a larger fraction of the Galactic Halo, extending it to the inner 10$^{\circ}$ of the Galaxy. The spatial ROIs are defined as circular concentric regions of 0.2$^{\circ}$ width each, centered at the GC, also excluding a $\pm$0.3$^{\circ}$ band in Galactic latitude. The ROIs widths were chosen so that each instrument has sufficiently angular resolution to resolve each ring separately and without much ``leakage'' onto the next ring.

%

\subsection{Analysis methodology}
\hfill 

Exclusion or sensitivity limits to DM annihilation can be derived by comparing the observed gamma-ray signal to the expected background. Following the previous studies, we use a 2D (energy and space) joint-likelihood method to derive our limits to secluded DM, where the comparisons between DM and background fluxes are performed in different energy and spatial intervals (or bins)~\cite{Abdallah:2016ygi,Viana:2019ucn}. The spatial bins, or ROIs, have been defined in previous section and the energy bins are defined for each instrument as: 70 logarithmically-spaced energy bins from $160$~GeV to $70$~TeV for H.E.S.S.~\cite{Abdallah:2016ygi}, 92 logarithmically-spaced bins from $0.02$~TeV to $100$~TeV for CTA~\cite{CTA_ScienceTDR}, and 40 logarithmically-spaced bins from $100$~GeV to $100$~TeV for SWGO~\cite{Viana:2019ucn}. 

The likelihood is calculated for each considered energy $E_{\gamma}$ as a product over the spatial ROIs (bins $i$) and the energy bins (bins $j$).  The 2D binned spatial and spectral likelihood formula is composed of a Poisson ``ON'' term (first term in Eq.~(\ref{eqn1})) and a Poisson ``OFF'' term (the second term in Eq.~(\ref{eqn1})), and it can be generally defined as:
\begin{equation}
\mathcal{L}_{ij}(s_{ij},b_{ij}| \mathcal{D}_{ij} ) = {\frac{(s_{ij}+b_{ij})^{N_{{\rm ON},ij}}}{N_{{\rm ON},ij}!}}e^{-(s_{ij}+b_{ij})}\times \frac{(s_{ij}' + \frac{b_{ij}}{\alpha_{i}})^{N_{{\rm OFF},i,j}}}{N_{{\rm OFF},ij}!}e^{-(s_{ij}' + \frac{b_{ij}}{\alpha_{i}})}
\label{eqn1}
\end{equation}
where the $\mathcal{D}_{ij}$ is the set containing the number of gamma-ray events observed in the signal region ($N_{{\rm ON},ij}$), in the control region ($N_{{\rm OFF},ij}$), and exposure ratio between signal and control regions $\alpha_i$. $s_{ij}$ is the number of predicted signal events in the bin ($i,j$), $b_{ij}$ the number of expected background events, and $s_{ij}'$ is the number of predicted signal events in the OFF region. The predicted signal is computed by folding the considered gamma-ray flux with the instrument response functions. The expected signal in a spatial bin $i$ and energy bin E$_{j}$ is given by

\begin{equation}
s_{ij} = T_{\rm obs} \int_{\Delta E_j} dE_{\gamma}^r \int_0^{\infty} dE_{\gamma}^t \frac{{\rm d}\Phi_{\gamma,i}(E_{\gamma}^t)}{{\rm d} E_{\gamma}^t} \times A_{\rm eff}(E_{\gamma}^t) \times {\rm PDF}(E_{\gamma}^t,E_{\gamma}^r)
\end{equation}   
where $T_{\rm obs}$ is the observation time, $E^t_{\gamma}$ is the true energy, $A_{\rm eff}$ is the effective collection area as function of the true energy, and ${\rm PDF}(E_{\gamma}^t,E_{\gamma}^r)$ is the representation of the energy resolution as the probability density function $P(E_{\gamma}^r|E_{\gamma}^t)$, of observing an event at the reconstructed energy $E_{\gamma}^r$ for a given true energy $E_{\gamma}^t$. 

The background observed counts of H.E.S.S. and CTA are estimated using the reflected background method. In the case of H.E.S.S., one OFF region was selected for each ON region, yielding an $\alpha_i = 1$, and the detected event's distribution for each ROI is extracted from Ref.~\cite{LefrancPhD}. Due to the larger field of view of CTA, we assumed five OFF regions for each ON region, i.e. $\alpha_i = 0.2$ for its analysis. In the case of a particle detector, such as SWGO, the background is typically estimated using a method called ``Direct Integration'' (see Ref.~\cite{HAWC_CRAB} for more details), which gives an extremely accurate measurement of the residual background (systematical uncertainty of a few parts in 10$^{-4}$). Hence, the OFF term of the likelihood function can be neglected, which is equivalent to assuming $\alpha_i \ll 1$. The CR  background counts of both CTA and SWGO are given by

\begin{equation}
N_{{\rm OFF},ij} = T_{\rm obs} \int_{\Delta \Omega_i} \int_{\Delta E_j}  dE d\Omega \frac{{\rm d} \Gamma^{\rm CR}}{{\rm d}E {\rm d}\Omega}
\end{equation}   
where ${\rm d} \Gamma^{\rm CR} / {\rm d}E {\rm d}\Omega$ is the differential residual background rate per steradian as a function of reconstructed energy, estimated through extensive Monte Carlo simulations and publicly available. The contamination of our signal and control regions by the Galactic Diffuse Emission (GDE) was not considered here. Up to this date, this emission has not been detected in the TeV energy range,  it was however shown in Ref.~\cite{Acharyya:2020sbj} that for the models obtained by extrapolating Fermi-LAT measurement to CTA energies, the GDE is expected to be at a sufficiently high level for CTA to detect it. It was shown that in an ON-OFF analysis that if properly modeled, the GDE degrades the CTA sensitivity by roughly a factor of 2~\cite{Lefranc:2015pza}. A thorough treatment of the different models and extrapolations from lower energies is beyond the scope of this paper, for a detailed study of the systematic effects of the GDE modeling in the case of CTA sensitivity to DM searches, see Ref.~\cite{Acharyya:2020sbj}.

The total  2D joint-likelihood function over the full energy range and all the spatial ROIs is the product of the individual likelihood functions over all bins $i$ and $j$,

\begin{equation}
\mathcal{L} (m_{\rm DM},\langle \sigma v \rangle ) = \prod_{ij} \mathcal{L}_{ij} \,,
\end{equation}
and constraints on $\langle \sigma v \rangle$ are obtained from the log-likelihood ratio test statistic given by $TS = - \ln(\mathcal{L}_0(m_{\rm DM},\langle \sigma v \rangle )/\mathcal{L}_{\rm max}(m_{\rm DM},\langle \sigma v \rangle) )$, where $\mathcal{L}_0$ is the null hypothesis (no DM model) likelihood and $\mathcal{L}_{\rm max}$ is the alternative hypothesis (with DM model) likelihood, evaluated at the value of the cross-section which maximizes the likelihood. Values of TS equal to 2.71 provides one-sided upper limits on $\langle \sigma v \rangle$ at a 95\% Confidence Level
(C.L.), assuming that the test statistic behaves as a ${\chi}^2$ distribution, as expected in the high statistic limit, with one degree of freedom.

\section{Results}
\label{sec:results}
\hfill 

\begin{figure}[!ht]
	\begin{center}	
		\includegraphics[width=0.49\linewidth]{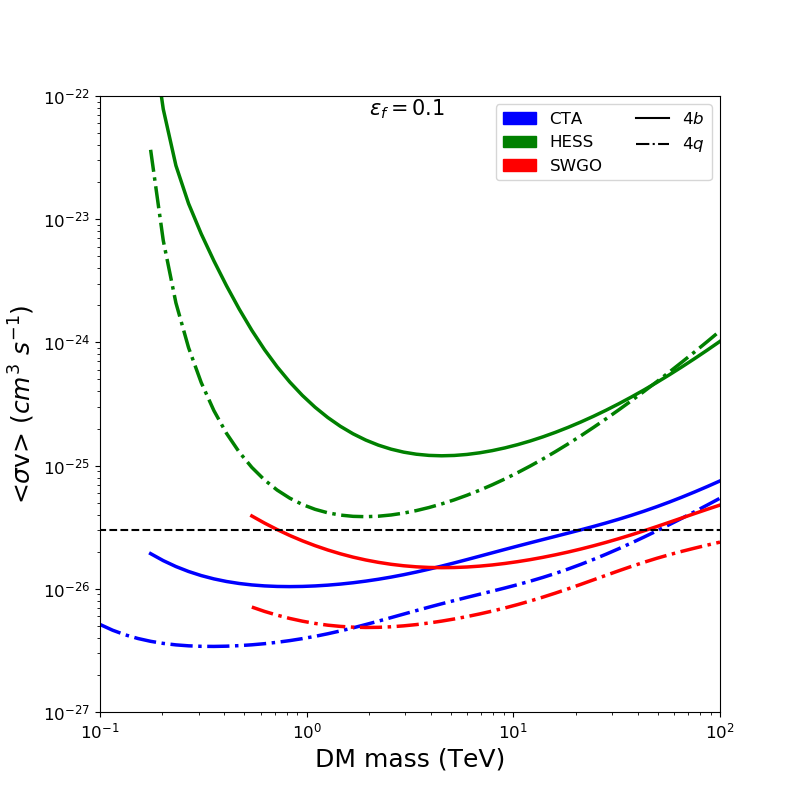}
		\includegraphics[width=0.49\linewidth]{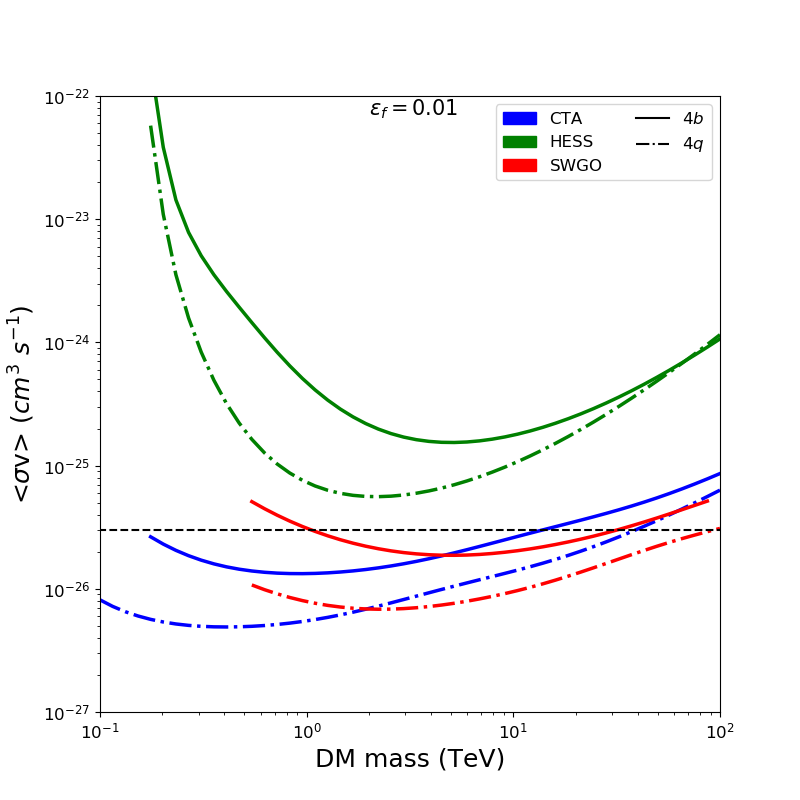}
	\end{center}
	\caption{Upper limit on the velocity-weighted cross-section with 95\% C.L. for DM self-annihilation into $V \to 2b$ (solid) and $V \to 2q$ (dash-dotted), with $\epsilon_f = 0.1$ (left) and $\epsilon_f =0.01$ (right), as a function of $m_{\rm DM}$, for the actual H.E.S.S. (green), for $T_{obs}=254$h, and the prospects for SWGO (red) and CTA (blue) observations of the GC halo, for 10 years and 500 hours of observation time, respectively. The nominal value of the thermal-relic cross-section \cite{PhysRevD.86.023506} is plotted as well (long-dashed black line).}
	\label{fig:DM_sens}
\end{figure}

In Figure~\ref{fig:DM_sens}, we present the limits for the $V\rightarrow 2b$ and $V\rightarrow 2q$ cases. Considering the $V\rightarrow 2q$ channel, H.E.S.S data exclude $\langle \sigma v \rangle \geq 4 \times 10^{-26}\,\, {\rm cm}^3 \, {\rm s}^{-1}$ for $m_{\rm DM}= 2$~TeV, while CTA and SWGO sharing similar sensitivities probe $\langle \sigma v \rangle \geq 5 \times 10^{-27}\,\, {\rm cm}^3\, {\rm s}^{-1}$ for the same mass. 
One can observe that for the DM annihilation into $4q$, the combination of CTA and SWGO will be able to probe the thermal cross-section in the entire mass range going from $100$~GeV to $100$~TeV. Comparing the left and right panels of Figure~\ref{fig:DM_sens}, we conclude that the change in the $\epsilon_f$ parameter from $0.1$ to $0.01$ yields no significant change, showing that our findings are robust. We remind the reader that $\epsilon_f =2 m_f/m_V$ accounts for the ratio between the fermion mass over the mediator mass. It simply assesses how heavy the mediator $V$ is compared to its decay products.

\begin{figure}[!ht]
	\begin{center}	
		\includegraphics[width=0.49\linewidth]{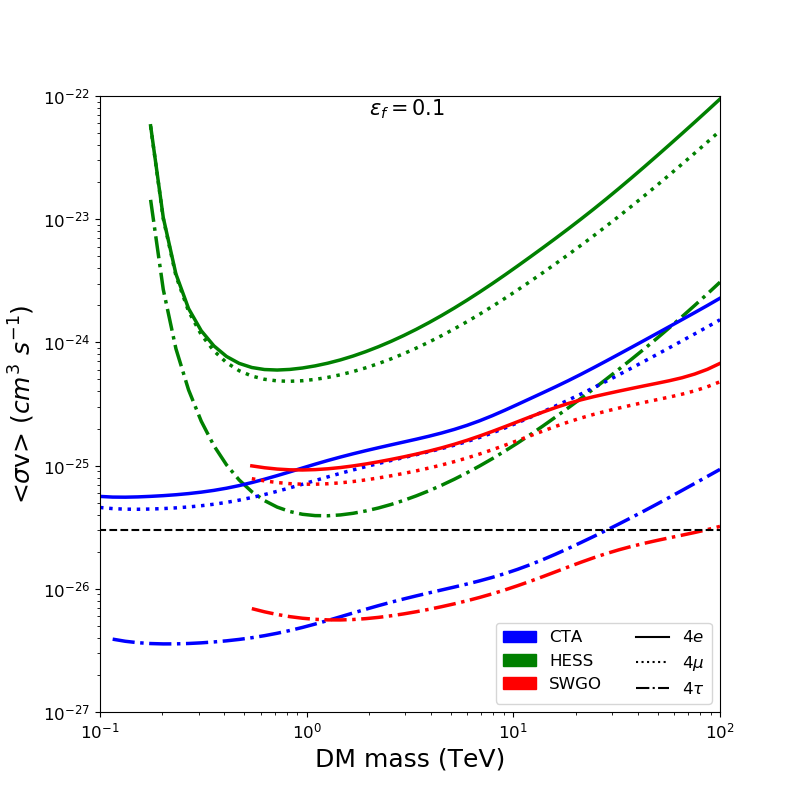}
		\includegraphics[width=0.49\linewidth]{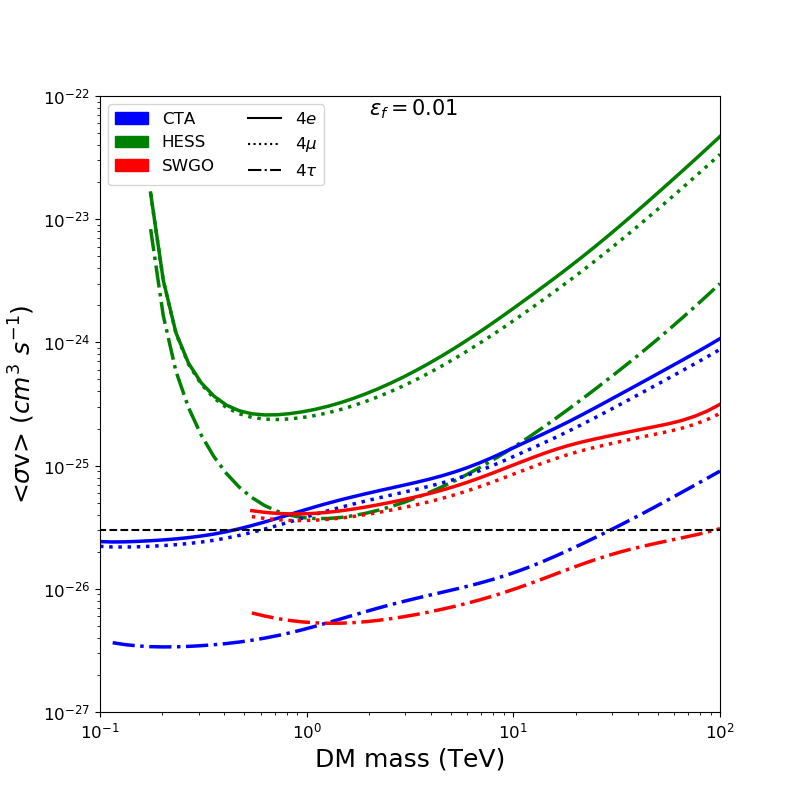}
	\end{center}
	\caption{Upper limit on the velocity-weighted cross-section with 95\% C.L. for DM self-annihilation into $V\to 2\tau$ (dashed), $V \to 2\mu$ (dotted) and $V \to 2e$ (solid), with $\epsilon_f = 0.1$ (left) and $\epsilon_f =0.01$ (right), as a function of $m_{\rm DM}$, for the actual H.E.S.S. (green), for $254$h of observation time, and the prospects for SWGO (red) and CTA (blue) observations of the GC halo, for 10 years and 500 hours of observation time, respectively. The nominal value of the thermal-relic cross-section \cite{PhysRevD.86.023506} is plotted as well (long-dashed black line).}
	\label{fig:DM_sens2}
\end{figure}

Moving to the leptonic annihilation states, in Figure~\ref{fig:DM_sens2}, we present in the left and right panels the current and projected upper limits on the annihilation cross-section for the $V\rightarrow 2e$, $V\rightarrow 2\mu$, $V\rightarrow 2\tau$ decay modes. As before, the only difference between the two panels is the value adopted for $\epsilon_f$. After looking carefully at the two plots, one can realize that when V decays into $\tau^+\tau^-$ the value of $\epsilon_f$ does not lead to any appreciable change in the upper limits. A similar conclusion occurred when the V decayed into hadrons (Figure~\ref{fig:DM_sens}). This can be understood because the hadronic and $\tau\tau$ final states yield energy spectra that look alike. For the cases when V decays into $e^+e^-$ and $\mu^+\mu^-$, the gamma-ray energy spectrum is very much dependent on the final state radiation that is sensitive to the energy of primary leptons, which in turn is determined by the mass of the mediator. Final state radiation processes are not as crucial for the $\tau$ lepton due to its relatively large mass. Therefore, as expected, the mass of the mediator is more relevant for the $e^+e^-$ and $\mu^+\mu^-$ decay modes. To illustrate the importance of CTA and SWGO, we will pick some benchmark points, one per energy decade, please see Table~\ref{tab:bench}. 

%
%
%
%
%
%
\renewcommand{\arraystretch}{1.3}
\begin{table}[!ht]
\centering
\begin{tabular}{ |c|c|c|c| } 
\hline
\multirow{2}{6em}{\bf\centering DM mass} & \multicolumn{3}{c|}{\bf Experiments }\\
\cline{2-4}
  & {\bf\color{DarkGreen} H.E.S.S. (current)} & {\bf\color{blue} CTA (prospects) } & {\bf\color{red} SWGO (prospects) } \\ 
 \hline
 \hline
$200$~GeV & $10^{-24}\,\, \rm cm^3\, s^{-1}$ & $ 5 \times 10^{-27}\,\, \rm cm^3\, s^{-1}$ & - \\ 
 \hline
$2$~TeV & $ 4 \times 10^{-26}\,\, {\rm cm}^3\, {\rm s}^{-1}$ & $ 7 \times 10^{-27}\,\, {\rm cm}^3\, {\rm s}^{-1}$ & $ 6 \times 10^{-27}\,\, {\rm cm}^3\, {\rm s}^{-1}$ \\ 
\hline
$10$~TeV & $  2 \times 10^{-25}\,\, {\rm cm}^3\, {\rm s}^{-1}$ & $ 10^{-26}\,\, {\rm cm}^3\, {\rm s}^{-1}$ & $  8\times 10^{-27}\,\, {\rm cm}^3\, {\rm s}^{-1}$ \\
\hline
$100$~TeV & $ 2 \times 10^{-24} \,\, {\rm cm}^3\, {\rm s}^{-1}$ & $ 10^{-25}\,\, {\rm cm}^3\, {\rm s}^{-1}$ & $ 2 \times 10^{-26}\,\, {\rm cm}^3\, {\rm s}^{-1}$ \\
\hline
\end{tabular}
\caption{Benchmark points for the upper limits on $\langle \sigma v \rangle$ (95\% C.L.) for ${\rm DM\, DM} \rightarrow V, V \rightarrow 4\tau$.}
\label{tab:bench}
\end{table}

\begin{figure}
    \centering
    \includegraphics[scale=0.5]{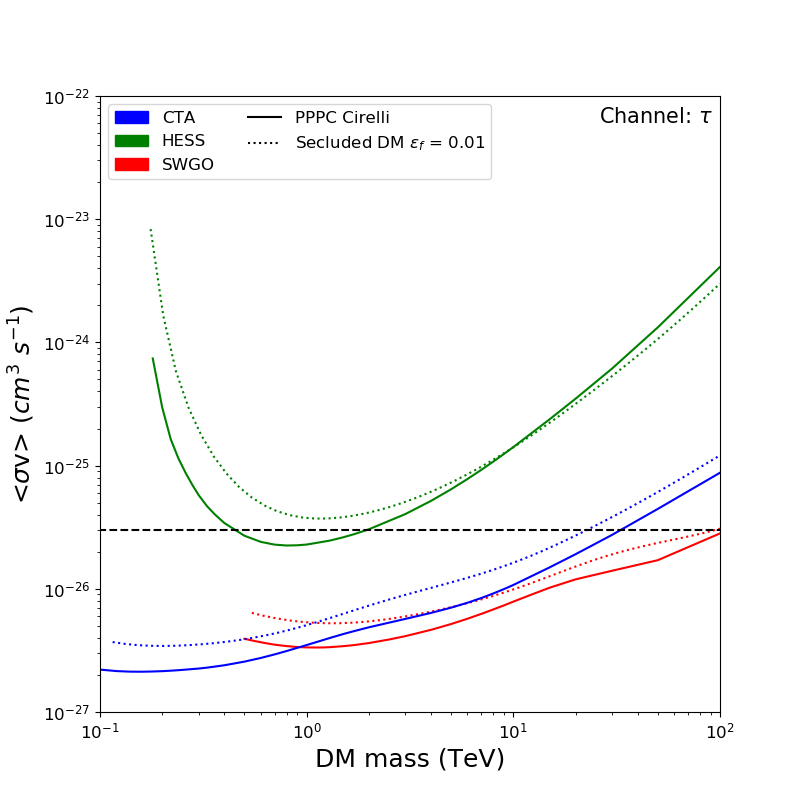}
    \caption{Comparison between a standard scenario (solid lines) and the secluded one (dotted lines) studied here. We choose the $\tau$ channel for an $\epsilon_f=0.01$. Please see the text for details.}
    \label{fig:comp}
\end{figure}

These benchmark points clearly show the importance of the CTA and SWGO telescopes. For $m_{DM} < 500$~GeV the gamma rays fall below the energy threshold of the SWGO instrument. Thus, CTA is the most constraining telescope at those energies. When we move to DM masses around $1$~TeV, CTA and SWGO share similar sensitivities, but as we explore DM masses in the multi-TeV range the SWGO telescope starts to prevail. This is in accordance with the better flux sensitivity of SWGO with respect to CTA in the multi-TeV energy range~\cite{SGSO_WP}. The same behavior is also shown in the other channels investigated here. For completeness, we also include a plot, please see Fig.~\ref{fig:comp}, comparing the secluded scenario (dotted lines) with the standard one (solid lines), for the $\tau$ channel with $\epsilon_f = 0.01$. As we may expect, the limits coming from the standard scenario are stronger than the secluded one, usually due to the broad spectrum generated by secluded models. 

\section{Conclusion}
\hfill 

In this study, we have calculated the sensitivity of current and future VHE gamma-ray observatories to secluded DM annihilations in the inner Galactic Halo. We have extended the reach of past searches for DM particles with masses beyond $10$~TeV, performing analysis over a mass range from $100$~GeV to $100$~TeV. We computed the exclusion limits using the H.E.S.S. dataset from 10 years of observations of the GC region with a total live time of $254$ hours. These results are compared to future sensitivities of the CTA and SWGO observatories, with 500 hours and 10 years of observations, respectively. 

Assuming an Einasto profile for the DM density distribution of the Milky Way, H.E.S.S. provides the most constraining exclusion limits from real data to secluded DM annihilations in the TeV mass range. However, it falls short of excluding the thermal relic cross-section value. Interestingly, with an additional $\sim$200 hours of exposure, H.E.S.S. could reach the thermal relic level for DM particles masses around 2~TeV in the $V\rightarrow 2q$ and $V\rightarrow 2\tau$ decay modes. The expected sensitivities of CTA and SWGO will be able to reach cross-sections below the thermal relic value for DM particles in the whole mass range between 100 GeV and 100 TeV in the $V\rightarrow 2q$ and $V\rightarrow 2\tau$ decay modes and between 100~GeV and $\sim$40~TeV in the $V\rightarrow 2b$ decay mode. It is important to emphasize that in this analysis we are considering that the GDE emission at TeV scale will be too faint to affect our results.  

In summary, in the next decades, we will have the opportunity to probe fully unexplored regions of parameter space of secluded DM models. Hence, these telescopes have the potential to observe a positive signal of DM annihilation that will help us unveil the fundamental nature of dark matter. Thus, it is quite clear the importance of new gamma-ray instruments. They will probe an entirely new region of parameter space, with a sensitivity much below the thermal annihilation cross-section.

\label{sec:conc}

\section*{Acknowledgements}
GNF and AV were sponsored by the S\~{a}o Paulo Research Foundation (FAPESP) through Grant No 2019/14893-3. GNF, AV, FSQ, and CS have been supported by the S\~{a}o Paulo Research Foundation (FAPESP) through Grant No 2015/15897-1. AV is supported by CNPq grant {\rm 314955/2021-6}. CS is supported by grant 2020/00320-9, S\~ao Paulo Research Foundation (FAPESP). FSQ acknowledges support from CNPq grants {\rm 303817/2018-6} and {\rm 421952/2018-0}, and ICTP-SAIFR FAPESP grant {\rm 2016/01343-7}, MEC, and UFRN. This work was supported by the Serrapilheira Institute (grant number Serra-1912-31613).\\

This paper has gone through an internal review by the CTA Consortium. This work was presented orally by C. Siqueira at the scientific event ``II Encontro de Primavera da SBF'' held from September 26 to 29, 2022 in Natal/RN.


\printbibliography

\end{document}